\documentstyle[11pt,paspconf]{article}

\begin{document}

\title{Chemical Abundances in the Carina dSph Galaxy}
\author{Tammy A. Smecker-Hane\altaffilmark{1}, Georgi I. Mandushev}
\affil{Dept. of Physics \& Astronomy, 4129 Reines Hall, University
      of California, Irvine, CA 92697-4575}
\author{James E. Hesser\altaffilmark{1}, Peter B. Stetson}
\affil{National Research Council of Canada, Herzberg Institute of Astrophys.,
    Dominion Astrophys. Obs., Victoria, BC V8X 4M6, Canada}
\author{G. S. Da Costa}
\affil{Research School of Astronomy \& Astrophysics, Australia National University, 
    Canberra, ACT 2611, Australia}
\author{Despina Hatzidimitriou}
\affil{Dept. of Physics, Univ. of Crete, Heraklion, Crete, Greece}

\altaffiltext{1}{Visiting Astronomer, Cerro Tololo Inter-American Observatory. 
CTIO is operated by AURA, Inc.\ under cooperative agreement with the NSF.}

\begin{abstract}
We report on the chemical abundances of stars in the
Carina dwarf spheroidal galaxy (dSph) derived from low-resolution
spectra. We have determined values of [Fe/H] for 52 stars from the
reduced equivalent width of the Ca II infrared triplet lines.
The Carina dSph has a mean metallicity of [Fe/H]$=-1.99 \pm 0.08$ 
and an intrinsic metallicity dispersion 0.25 dex (1$\sigma$).
By directly determining the chemical abundances of
Carina stars through spectroscopy, we can overcome the 
age--metallicity degeneracy inherent in color-magnitude diagrams (CMDs)
and determine its star-formation history with unprecedented
accuracy.
\end{abstract}


\vspace{-1ex}
\section{Introduction}

The dSphs in the Local Group offer a unique opportunity for
testing our understanding of galaxy evolution. New CMDs obtained
for dSphs show that most have
undergone complex evolution.  They formed stars over
{\it many Gyr} rather than on a dynamical timescale (few $ \times 
10^8$ yr) despite the fact that they are now devoid of 
gas and their low total masses ($10^7$ to $10^9 M_\odot$) 
make them highly susceptible to supernova-driven 
galactic winds. (For a recent review, see Smecker-Hane 1997.) 
We have undertaken as series of studies involving 
photometry and spectroscopy of stars in the Carina and 
Fornax dSphs to determine their star-formation and chemical 
evolution histories.  Here we report on our spectroscopy of
Carina dSphs stars and their derived metallicities.

\section{Observations and Data Analysis}

We used our photometric survey of bright stars centered on the
Carina dSph (Smecker-Hane, et al. 1994) to identify red giants
that were probable members of the galaxy.
At a Galactic latitude of $b=-22^\circ$, the Carina fields 
are contaminated with a significant number of Galactic foreground 
stars.  Although members cannot be positively identified by their
location in the CMD, they are easily identified by
their large radial velocities (223 km/sec). For our spectroscopic 
sample, we randomly chose stars from the CMD that had magnitudes 
within 1.5 mag of the tip of the red giant branch (RGB), 
(B-I) colors within $\pm 0.14$ of the mean color of 
the RGB, and distances $<11$ arcmin (1 core radius) from the 
center of the galaxy.

We obtained spectra of these stars with the Cerro Tololo
Inter-American Observatory 4.0-meter telescope and the ARGUS
multi-fiber system in November, 1997. The ARGUS system allows
simultaneous observation of 24 stars and 24 sky positions. 
Total exposure times were 3.5 to 5 hours.
Images were reduced using the standard CCD reduction package
in the Image Reduction and Analysis Facility (IRAF), and the
spectra were extracted and calibrated with the ARGUS package.
The most difficult part of the data reduction was the sky
subtraction because the Ca II lines fall among 
bright night sky emission lines. The fluxes in the sky lines 
were typically 5 times higher than the stellar continuum fluxes in
the region of the the Ca II lines. The final stellar spectra 
have a dispersion of 0.83 \AA/pix, and a typical signal-to-noise 
ratio of 17 per pixel. Examples are shown in Figure~1.

\begin{figure}
\plotfiddle{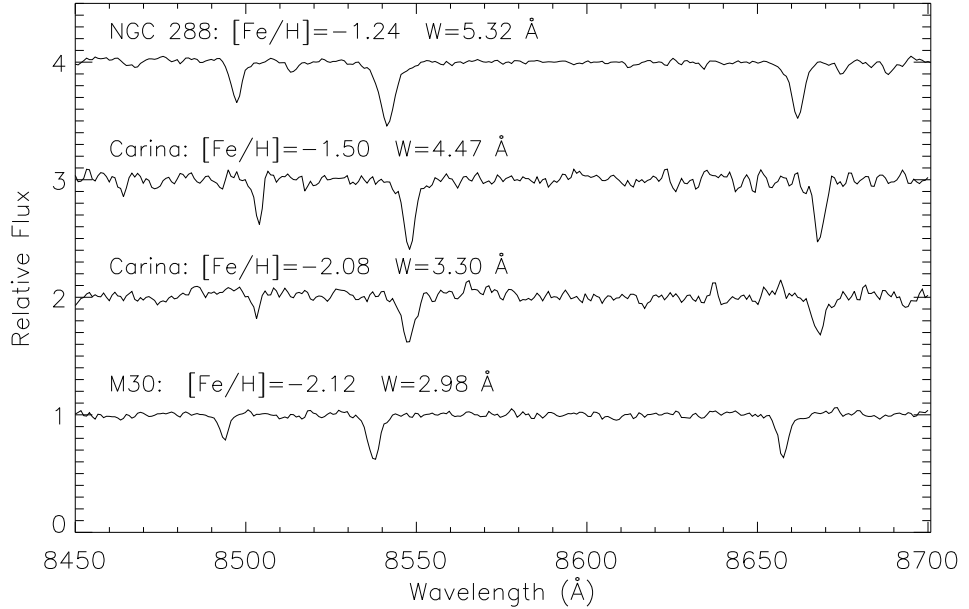}{7cm}{0.}{70.}{70.}{-210}{-285}
\vspace{0.7cm} \caption{Spectra of Carina dSph and globular cluster stars. 
The Ca~II lines at 8542 and 8662 \AA ~are used to derive metallicities.}
\end{figure}

\section{Chemical Abundances}

We derived metallicities, [Fe/H], from the reduced equivalent
width, $W^\prime$, of the Ca II infrared triplet lines in each
spectrum. This technique has been empirically calibrated by
Rutledge, et al. (1997) using observations of numerous Galactic
globular cluster stars, and interested readers are referred to it
for details. In brief, the sum of the equivalent widths of the two
strongest lines of the Ca II triplet at 8542 and 8662
\AA, $W$, shows a linear dependence with [Fe/H] when the effect of
effective temperature and gravity are removed to first order. The
later is effectively done by forming the reduced equivalent
width for each star, $W^\prime \equiv W + 0.62
(V-V_{HB})$, where $V$ is the magnitude of the star and $V_{HB}$ is
the magnitude of the horizontal branch. The equivalent width for each
Ca II line was determined by fitting a linear continuum plus 
Gaussian-shaped absorption feature in the standard wavelength regions
defined by Armandroff \& Da Costa (1991). The adopted
calibration equation, [Fe/H] $= -2.66 + 0.42 W^\prime$, 
from Rutledge, et al. (1997) gives metallicities based
on the scale defined by the high-dispersion spectroscopic work of
Carretta \& Gratton (1997).

Figure 2 shows a histogram of the derived metallicities for 52
radial velocity members of Carina and a Gaussian fit to the histogram. 
The average metallicity is [Fe/H] $ = -1.99 \pm 0.08$ (including
the calibration zeropoint error in the uncertainty) and the observed 
dispersion is $\sigma = 0.30 \pm 0.02$ dex.  The error in each 
derived metallicity is set primarily by 
the sky flux and random fiber-star positioning errors.
Thus the [Fe/H] measurement errors are, to first order,
independent of the magnitude and metallicity of the stars in
the limited magnitude range over which we are observing.
Subtracting in quadrature the average measurement error of 
0.15 dex and 0.02 dex uncertainty in the slope of the calibration 
from the observed dispersion reveals an intrinsic 
metallicity dispersion of $\sigma = 0.25$ dex (1.0 dex full 
width) for the Carina dSph.

\begin{figure}
\plotfiddle{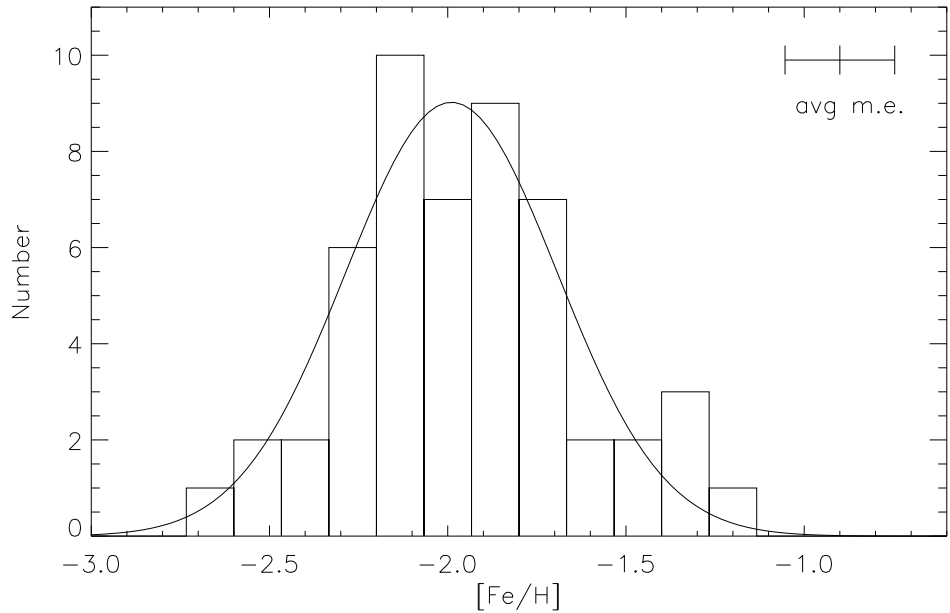}{7cm}{0.}{70.}{70.}{-210}{-280}
\vspace{0.5cm} \caption{The metallicity histogram for 52 members
of the Carina dSph galaxy.  The average 1--$\sigma$ measurement 
error of 0.15 dex is shown in the upper right-hand corner.}
\end{figure}

Two notable calibration issues deserve discussion.
First, the $W^\prime$--[Fe/H] calibration is strictly valid for old stars
although Armandroff \& Da Costa (1991) note no 
strong age dependence. Second, the calibrating Galactic globular 
clusters exhibit a relatively well-defined relationship in [Ca/Fe]
versus [Fe/H] that reflects the history of enrichment from 
Type~Ia and Type~II supernovae in the Galaxy. However dwarf galaxies may have
entirely different star-formation histories and may not share the
same [Ca/Fe]-[Fe/H] relationship 
(see Smecker-Hane \& McWilliam in these proceedings for a 
discussion of abundance ratios in the Sagittarius dSph).
Therefore, we are embarking on a series of observations to calibrate
$W^\prime$ to yield [Ca/H] and to explore its sensitivity to age.

\section{Discussion}

The derived mean metallicity and intrinsic metallicity dispersion
roughly agree with our initial analysis of the Carina CMD
(Smecker-Hane 1997) although the metallicity dispersion derived 
from spectroscopy is larger. A full analysis of the CMD is
underway to determine Carina's star-formation history. 
Carina's RGB is very narrow in color because of the degeneracy of 
age and metallicity.  A wide range in age ($\sim$ 2 to 14 Gyr)
counteracts a 1.0 dex spread in metallicity in such a way that the 
younger RGB stars are more metal rich and hence have approximately 
the same colors as the older, more metal-poor, RGB stars. 
Thus a galaxy with a narrow RGB may indeed have a complex stellar
population.  An
interesting contrast to the Carina dSph is the Leo I dSph.
From analysis of its CMD, Gallart, et al. (1999) conclude that
its wide RGB may be a result of its wide range in age 
($\sim$ 1 to 10 -- 15 Gyr) and its internal metallicity dispersion
may be very small.  In conclusion, the dSphs exhibit a
very wide variety of star-formation and chemical evolution
histories, and we can learn much about the physical mechanisms
that control their evolution by studying them in detail.

\acknowledgments We thank the CTIO TAC for granting us
time to make these observations, the CTIO staff for their
outstanding support, and the NSF for financial support through grant
AST-961946 to TSH.


\begin{references}
\reference Armandroff, T. E. \& Da Costa, G. S. 1991, AJ, 101,
     1329
\reference Carretta, E. \& Gratton, R. G. 1997, A\&AS, 121, 95
\reference Gallart, C., et al. 1999, ApJ, 514, 665
\reference Rutledge, G. A., Hesser, J. E., Stetson, P.
      B. 1997,PASP, 109, 907 \reference Smecker, T. A. 1997,
      in {\it Star Formation Near and
      Far. Seventh Astrophysics Conference at College Park Maryland},
      ed. S. S. Holt and L. G. Mundy (AIP Press: New York), 571
\reference Smecker-Hane, T. A., Stetson, P. B., Hesser, J. E. \&
       Lehnert, M. D. 1994, AJ, 108, 507
\end{references}
\end{document}